\documentclass[twocolumn]{aastex631}

\usepackage{balance}

\newcommand{\linemake}{\texttt{linemake}}

\begin{document}

\title{\linemake:\ an Atomic and Molecular Line List Generator}

\author[0000-0003-4479-1265]{Vinicius M. Placco}
\affiliation{Community Science and Data Center/NSF’s NOIRLab, 950 N. Cherry Ave., Tucson, AZ 85719, USA}

\author[0000-0002-3456-5929]{Christopher Sneden}
\affiliation{Department of Astronomy and McDonald Observatory,
The University of Texas, Austin, TX 78712, USA}

\author[0000-0001-5107-8930]{Ian U. Roederer}
\affiliation{Department of Astronomy, University of Michigan, 1085 S. University Ave., Ann Arbor, MI 48109, USA}
\affiliation{Joint Institute for Nuclear Astrophysics -- Center for the Evolution of the Elements (JINA-CEE), USA}

\author[0000-0001-5579-9233]{James E. Lawler}
\affiliation{Department of Physics, University of Wisconsin-Madison,
1150 University Ave., Madison, WI 53706}

\author[0000-0001-8582-0910]{Elizabeth A. Den Hartog}
\affiliation{Department of Physics, University of Wisconsin-Madison,
1150 University Ave., Madison, WI 53706}

\author[0000-0001-5541-6087]{Neda Hejazi}
\affiliation{Department of Physics and Astronomy, Georgia State University, Atlanta, GA 30302, USA}

\author[0000-0002-0475-3662]{Zachary Maas}
\affiliation{Department of Astronomy and McDonald Observatory,
The University of Texas, Austin, TX 78712, USA}

\author[0000-0002-1255-396X]{Peter Bernath}
\affiliation{Department of Chemistry and Biochemistry, Old Dominion University, Norfolk, VA 23529, USA}
\affiliation{Department of Physics, Old Dominion University, Norfolk, VA 23529, USA}

\begin{abstract}

In this research note, we present \linemake, an open-source atomic and molecular line list generator. Rather than a replacement for a number of well-established atomic and molecular spectral databases, \linemake\ aims to be a lightweight, easy-to-use tool to generate formatted and curated lists suitable for spectral synthesis work.  We encourage users of \linemake\ to understand the sources of their transition data and cite them as appropriate in published work. We provide the code, line database, and an extensive list of literature references in a \texttt{GitHub} repository (\href{https://github.com/vmplacco/linemake}{\texttt{https://github.com/vmplacco/linemake}}), which will be updated regularly as new data become available.

\end{abstract}


\section{Introduction} \label{sec:intro}

Stellar and Galactic chemical evolution can be illuminated by abundance analyses, but such studies must strive for abundance accuracy.  This effort depends on many factors, but none is more crucial than access to trustworthy basic atomic and molecular transition data.  Several excellent large transition databases exist, the most well-known being \texttt{VALD}\footnote{\href{http://vald.astro.uu.se/}{\texttt{http://vald.astro.uu.se/}}}.  Here we introduce the utility code \linemake, created to generate synthetic spectrum input line lists by merging Kurucz\footnote{\href{http://kurucz.harvard.edu/linelists.html}{\texttt{http://kurucz.harvard.edu/linelists.html}}} \citep{kurucz2011} atomic/molecular line compendium information with updated and very accurate transition probabilities, hyperfine structure (HFS), and isotopic substructure data. The atomic data included in \linemake\ are primarily those that have been published by the University of Wisconsin atomic physics group \citep[e.g.][among others]{lawler2009} and the molecular data are from the Old Dominion University molecular physics group \citep{bernath2020}. The Wisconsin data now include nearly all Fe- and lanthanide-group elements, and recently a study of \ion{Ca}{1} initiates work on lighter elements.

With increased interest in low-temperature stars (spectral types mid-K and cooler) it has become attractive to include molecules such as TiO and H$_2$O in \linemake. These molecules have substantial numbers of transitions available in large databases but are not easy to incorporate in synthetic spectrum line lists. They are included in \linemake\, with choices that keep the number of transitions to a manageable size. \linemake\ does not try to compete with the \texttt{VALD} resource or other more comprehensive line compendia. Our effort is to produce a carefully controlled line list of easily referenced sources from laboratory physics. In this document, we outline \linemake\ and point to the sources of its data.

%[VINI]: this was added to the REAME GitHub file
%The choices of which lines of which species to include in \linemake\ have often been driven by the authors' own spectroscopic interests (e.g., note the large number of entries for transitions of neutron-capture elements that can only be detected in vacuum-UV spectroscopy). However, we would welcome hearing from users who can suggest other strongly-sourced species (with recent reliable lab/theory results) that might be added to our database. 

\section{About \linemake}

\begin{figure*}
\epsscale{1.175}
\plotone{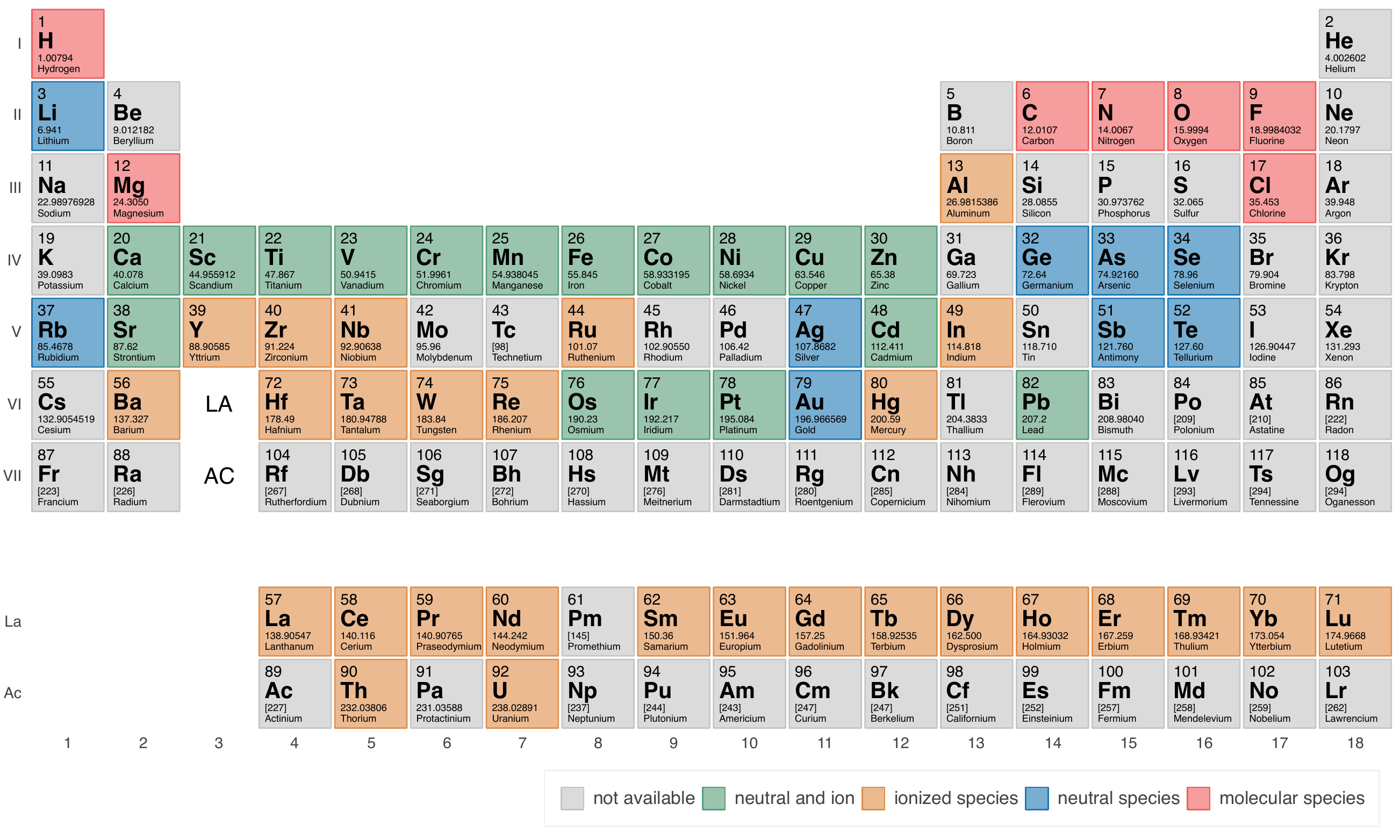}
\caption{Elements with \textbf{curated} transitions currently available in \linemake. The most up-to-date version of this periodic table is available at the \texttt{\href{https://github.com/vmplacco/linemake}{GitHub}} repository.}
\label{ptable}
\end{figure*}
%\href{http://vmplacco.github.io/files/linemake_ptable.html}{vmplacco.github.io/files/linemake\_ptable.html}
%\subsection{Description of the code}

\linemake\ produces synthesis line lists compatible with those needed by the line analysis code \texttt{MOOG}\footnote{\href{https://www.as.utexas.edu/~chris/moog.html}{\texttt{https://www.as.utexas.edu/\~{}chris/moog.html}}} \citep{sneden1973}.  It starts with the Kurucz compendium and then substitutes or supplements these lists with atomic data from the Wisconsin group and molecular data from the Old Dominion group. \linemake\ is written in standard \texttt{FORTRAN}, and it can be compiled and executed on all \texttt{UNIX/Linux}-based operating systems.

\linemake\ is primarily aimed for spectroscopic studies of stars cooler than B spectral type.  Some limitations in the available line lists have been imposed in order to keep the resulting synthesis lists to sensible sizes.  The ionization states available are neutral and first ion. The maximum lower excitation energy of any line is 7.5~eV, except for some transitions of the light elements H, C, N, O, Mg, Al, Si, P, and S; higher excitation transitions are available for these elements. Additionally, for \ion{Fe}{2}, 8.5~eV is the maximum lower energy. Obvious warnings should be given about the output line lists:\ we believe that they are correct, but there is no substitute for users having a close look to assure themselves of the quality of these lists. If a user is uncertain about an output line list, the individual files for different species can be examined easily. These species files are also useful in simply making laboratory-based reliable transition lists for individual (mostly atomic) species.

%\subsection{Using \linemake}

%\linemake\ is written in standard FORTRAN, and can be compiled/executed on all unix/linux operating systems.
A few ``local rules'' apply when using the code. First and most importantly, in almost all cases in which an atomic (and more often molecular) species is represented by multiple isotopes in its transitions, \linemake\ presents the lists with isotopic identification but no assumptions about isotopic ratios. For a given transition, the total $gf$ value for each isotope is the same. When applying spectrum synthesis codes like \texttt{MOOG} to the line lists generated by \linemake, users must set the desired isotopic fractions in parameter files. Secondly, \linemake\ catalogs some Fe-group hyperfine substructure patterns from the Kurucz database when no recent laboratory studies are available. These can be included in output files at the user's option, but they will probably add many extra transitions into the synthesis line lists, and the pedigrees of these are not guaranteed. Users are cautioned to be careful about including them and are encouraged to test them with stellar syntheses.
% [VINI] I'd suggest leaving this out. In the (not so distant) future, I want to add a bash wrapper to linemake to help with this issue.
%[VINI]: this was added to the REAME GitHub file
%Finally, \linemake\ will not work when the requested beginning and ending wavelengths bridge the divide between two files of atomic line data, each of which covers 1000\,{\AA}. The simple work-around is split the wavelength request and run the code twice. 

\section{Brief description of the database}

Here we briefly discuss the atomic and molecular data sources. 
%The Fe-group atomic species are considered first, followed by neutron-capture species, selected light elements, and molecular species. 
The detailed description for each category, including literature references and decisions made to maximize the utility of line lists for high-resolution spectroscopic studies, are given in the \texttt{\href{https://github.com/vmplacco/linemake/blob/master/README.md}{README.md}} file of the \texttt{\href{https://github.com/vmplacco/linemake}{GitHub}} repository. Figure~\ref{ptable} summarizes the transition availability in the \linemake\ database, and its most up-to-date version can also be found in the \texttt{\href{https://github.com/vmplacco/linemake/blob/master/README.md}{README.md}} file. %(\href{https://github.com/vmplacco/linemake}{https://github.com/vmplacco/linemake}).

\subsection{Atomic Species:\ Fe-group Elements}

The Fe group is defined here as those elements with $21 \leq Z \leq 30$. 
Significant HFS  usually is a feature of transitions of odd-$Z$ elements. If the \texttt{\href{https://github.com/vmplacco/linemake/blob/master/README.md}{README.md}} description indicates that a species includes HFS, then line lists have full HFS substructure patterns when they are available from laboratory studies. Typically, these are known for many of the lines with laboratory transition probabilities. \linemake\ manages these data internally, but users should examine the output synthesis lists to understand whether HFS patterns have been included in any transition of interest. Information about isotopic substructure is scarce. This is largely an issue for even-$Z$ elements, as odd-$Z$ elements have few (often just one) naturally-occurring isotopes. Moreover, for Fe-group elements, usually one isotope dominates the solar system abundance, and other isotopes can safely be ignored. Exceptions are Ti and Ni, and users should account for isotopic effects on transitions in the red and infrared spectral regions.
%[WHAT IS MEANT BY ``CONSIDER''?] -> perhaps "account for"?

\subsection{Atomic Species:\ Neutron-Capture Elements}

Neutron-capture elements are defined as those with $Z > 30$. 
For lanthanide elements and Hf ($ 57\leq Z \leq 72$), the Wisconsin group results dominate, but for lighter and heavier elements other sources are included if they have been subject to recent assessment for reliability. In this case,  \texttt{NIST} refers to the National Institute of Standards and Technology's Atomic Spectra Database.\footnote{\href{https://physics.nist.gov/PhysRefData/ASD/lines\_form.html}{\texttt{https://physics.nist.gov/PhysRefData/ASD/lines\_form.html}}}

\subsection{Atomic Species:\ Other Elements}

Light elements ($Z \leq 20$) have often not had recent extensive laboratory transition probability studies. Some progress, both in laboratory and theoretical, has been made, and a few selected elements are included in the \linemake\ database. 

\subsection{Molecular (mostly diatomic) Species}

Line data for molecular species have improved considerably recently. 
%Initially, we included molecules familiar to anyone doing optical-region spectroscopy of G and K stars. 
Motivated by an interest in M-type stars, which are often involved in exoplanet radial velocity or atmospheric transmission studies, we have included molecules that only appear in stellar spectra of very cool stars. The data sources for many molecules are laboratory studies from Peter Bernath's group (\mbox{\texttt{MoLLIST}}\footnote{\href{http://bernath.uwaterloo.ca/molecularlists.php}{\texttt{http://bernath.uwaterloo.ca/molecularlists.php}}}). For others we have translated into \texttt{MOOG} format the line information from two prominent molecular physics groups: \texttt{HITRAN}\footnote{\href{https://hitran.org/lbl/}{\texttt{https://hitran.org/lbl/}}} and \texttt{EXOMOL}\footnote{\href{http://www.exomol.com/data/molecules/}{\texttt{http://www.exomol.com/data/molecules/}}}.

\section{Conclusions}

In this note, we have briefly described the main features of the \linemake\ code. This discussion is not intended to be exhaustive, but we hope it may serve as a helpful guideline for potential users. The \texttt{\href{https://github.com/vmplacco/linemake}{GitHub}} database will continue to be updated with the most recent laboratory and theoretical work, and input from the user community is certainly welcome.

%\begin{acknowledgments}
%
%The work of V.M.P. is supported by NOIRLab, which is managed by the Association of Universities for Research in Astronomy (AURA) under a cooperative agreement with the National Science Foundation.
%
%I.U.R. acknowledges financial support from grants AST 16-13536, AST-1815403, and PHY 14-30152 (Physics Frontier Center/JINA-CEE) awarded by the NSF.
%
%\end{acknowledgments}

\onecolumngrid

\bibliography{linemake.bbl}{}
\bibliographystyle{aasjournal}

\end{document}